\documentclass[prl,twocolumn,showpacs]{revtex4}
\usepackage{amsmath,amssymb,color,bm,graphicx}
\usepackage[colorlinks=true,linkcolor=blue,citecolor=blue,urlcolor=red]{hyperref}

\newcommand{\omnu}{\omega_{\bar\nu}}
\newcommand{\ome}{\omega_{e}}
\newcommand{\sla}{\! \not \!}

\newcommand{\Img}{\Im{\rm m}}

\newcommand{\beq}{\begin{equation}}
\newcommand{\eeq}{\end{equation}}
\newcommand{\be}{\begin{equation}}
\newcommand{\ee}{\end{equation}}
\newcommand{\ba}{\begin{array}}
\newcommand{\ea}{\end{array}}
\newcommand{\bc}{\begin{center}}
\newcommand{\ec}{\end{center}}
\newcommand{\bea}{\begin{eqnarray}}
\newcommand{\eea}{\end{eqnarray}}
\newcommand{\bi}{\begin{itemize}}  
\newcommand{\ei}{\end{itemize}}
\newcommand{\ben}{\begin{enumerate}} 
\newcommand{\een}{\end{enumerate}}

\renewcommand{\Re}{{\rm Re}\,}

\usepackage[normalem]{ulem}  

\definecolor{red}{rgb}{0.8,0,0}
\definecolor{violet}{rgb}{0.4,0,0.4}
\definecolor{green}{rgb}{0,0.5,0.0}
\definecolor{navy}{rgb}{0.0,0.0,0.6}
\definecolor{orange}{rgb}{0.8,0.2,0.0}
\definecolor{blue}{rgb}{0.3,0.0,0.8}

\begin{document}

\title{Short-Range Correlations and Urca Process in Neutron Stars}

\author{Armen Sedrakian}%
\affiliation{Frankfurt Institute for Advanced Studies,
D-60438 Frankfurt am Main, Germany\\ and Institute of Theoretical Physics,
University of Wroclaw, 50-204 Wroclaw, Poland}  
\email{sedrakian@fias.uni-frankfurt.de}

\begin{abstract}
  Recent measurements of high-momentum correlated neutron-proton pairs
  at JLab suggest that the dense nucleonic component of the compact
  stars contains a fraction of high-momentum neutron-proton pairs that
  is not accounted for in the familiar Fermi-liquid theory of the
  neutron-proton fluid mixture. We compute the rate of the Urca
  process in compact stars taking into account the non-Fermi liquid
  contributions to the proton's spectral widths induced by short-range
  correlations. The Urca rate differs strongly from the Fermi-liquid
  prediction at low temperatures, in particular, the high threshold on
  the proton fraction precluding the Urca process in neutron stars is
  replaced by a smooth increase with the proton fraction. This observation
  may have a profound impact on the theories of cooling of compact
  stars.
\end{abstract}

\date{20 June 2014} 

\pacs{97.60.Jd}

\maketitle

{\it Introduction---}The Landau Fermi-liquid theory forms the basis of
semiphenomenological description of many low-temperature strongly
correlated fermionic systems, including liquid $^3$He in its normal
state, electrons in metals, nuclear matter, etc.  It assumes that slow
enough switching of interaction adiabatically evolves the low-lying
states, therefore the strongly interacting regime features a sharp Fermi
surface, whereas the excitations are weakly interacting
quasiparticles~\cite{LeggettBook}.
More recent research discovered that the low-energy thermodynamics and
transport of a wide class of condensed matter systems, including doped
metals, heavy fermions, $d$-$f$ electrons in metals, quantum dots, etc.
do not agree with the predictions of Fermi-liquid
theory~\cite{2001RvMP...73..797S,2002PhR...361..267V}. The
non-Fermi-liquid (nFL) behavior of these systems is often associated
with the breakdown of the concepts of well-defined quasiparticles,
Fermi surfaces, or both.

There exists now sufficient evidence that the proton component of the
high-density matter in neutron stars is an example of nuclear nFL.
Recent high-energy electron-scattering measurements at
JLab~\cite{2014Sci...346..614H} which used medium to heavy nuclear
targets show that in neutron-rich nuclei, short-range interactions
between the fermions form correlated high-momentum neutron-proton
pairs. This, in turn, implies that protons can occupy momentum states
that are much greater than their Fermi momentum, i.~e., the concept of
a sharp Fermi surface does not apply. Theoretical interpretations are
based on short-range correlations (SRCs) in the isospin singlet
channel and led to phenomenological modeling of the proton momentum
distribution tails~\cite{PhysRevC.89.034305}.  Microscopic
calculations of isospin asymmetrical nuclear systems confirm, at least
qualitatively, the presence and magnitude of SRCs in the momentum
distributions of infinite nuclear matter and in
nuclei~\cite{Frick2005,2014PhRvC..89d4303R}.

Neutrino emissivities of neutron stars play a central role in the
modeling their thermal evolution.  Neutron star cores that are
composed of neutrons ($n$), protons ($p$) and electrons ($e$) and cool
predominantly by emission of electron neutrinos $\nu_e$ and
antineutrinos $\bar\nu_e$ via the direct Urca
(dUrca)~\cite{Lattimer1991} and modified Urca (mUrca)
processes~\cite{Chiu1964,Friman1979}:
\bea  n &\to & p+e+\bar\nu_e,\\
 N+n&\to & N+p+e+\bar\nu_e,\quad N\in p, n
\eea
up to the onset of the superfluidity of the nucleonic components,
after which these processes are suppressed by the pairing gap in the
spectra of neutrons and protons, for a review
see~\cite{Sedrakian2019}. The small fraction of muons is neglected
here.

The mUrca process has been studied in pioneering
work~\cite{Chiu1964,Friman1979} as the 
dominant cooling process in neutron stars.
(We do not consider here the interesting possibility that the neutron
star cores consist of hypernuclear or deconfined quark matter.)
The more efficient Urca process was later discovered by Lattimer {\it et
al}.~\cite{Lattimer1991} to occur in the neutron-proton-electron
Fermi-liquid mixture above the threshold $x_p = n_p/n \ge 0.11$, where
$n_{\tau}$ $\tau \in n,p$ denote the particle numbers of neutrons and protons
and $n\equiv n_n+n_p$~\cite{Lattimer1991}. This threshold is reflected
in the step function in the Fermi-liquid dUrca neutrino emissivity
\bea\label{eq:epsilon_0}
\epsilon_{\rm Urca} = \epsilon_0\, \theta( \vert
p_{Fe}+p_{Fp}-p_{Fn}\vert),
\eea
$p_{Fi}$ being the Fermi momentum of particle $i \in n,p,e$. If
kinematically allowed the emissivity of the dUrca process is given
by~\cite{Lattimer1991}
\bea 
\epsilon_0 = \frac{457\pi}{10080} g_A^2G^2 p_{Fe}m_p^*m_n^*T^6,
\eea
where $m_i^*$ is the effective mass of particle $i$, $G$ is the weak
coupling constant, $g_A$ is the axial vector coupling constant, $T$ is
the temperature~\footnote{For the sake of simplicity we drop the
  numerically subdominant vector current neutrino emission.}. The
``triangle relation'' enforced by this step function requires proton
concentration $x_p\ge 0.11$ to fulfill the momentum and energy
conservation.  The nFL nature of the proton component argued above may
mitigate this constraint~\cite{2008IJMPA..23.2991F, McGauley}. To
date, the SRC-induced modification of the dUrca process was assessed
in terms of correction factors built from the momentum distribution of
protons in dense matter~\cite{2008IJMPA..23.2991F, McGauley}.  Early
investigations of the mUrca process~\cite{Chiu1964,Friman1979} were
later enhanced by more sophisticated calculations that incorporated
the Brueckner $G$-matrix~\cite{Shternin2018} or in-medium soft pion
exchange~\cite{Voskresensky1990}, along with wave-function
renormalization factors in the nucleon density of
states~\cite{Dong2016}. Additionally, Shternin {\it et
al.}~\cite{Shternin2018} showed that refining the intermediate-state
propagator significantly increases mUrca rates near the dUrca
threshold. These advancements are consistent with the Fermi-liquid
paradigm, as both asymptotic and intermediate states are well-defined
Landau quasiparticles.

In this Letter we compute the Urca process microscopically in
terms of the nFL spectral function of the proton component. By doing
so we establish direct contact with the fundamental microscopic
quantities -- the self-energies of the system, which can be computed in
a controlled many-body theory. Physically, the tensor interactions
between the minority component of protons and the majority component
of neutrons leads to the nFL behavior, which is understood herein as
the need to account for proton quasiparticle energy-momentum dependent
width, i.e., the full spectral function.  We show that in the
quasiparticle limit for proton spectral function we naturally recover
the standard Fermi-liquid result~\cite{Lattimer1991}. We also argue
that the dichotomy of the cooling scenarios of neutron stars (fast vs
slow) triggered by the dUrca process in massive stars will be smoothed
out if the nFL rates are used instead, i.~e., the paradigm of fast
cooling of (only) massive neutron stars via the dUrca process should
be abandoned and replaced by a smooth transition from slow to fast
cooling scenarios. Before proceeding, let us mention that the finite
width effects have been considered in the context of neutrino
bremsstrahlung in the pioneering work by Raffelt {\it et
al.}~\cite{Raffelt1996,Raffelt1999} and have been later put on more
formal grounds in the real-time Green's functions formalism
in~\cite{Sedrakian1999,Sedrakian2000}.

{\it Formalism---}The antineutrino emissivity is given by~\cite{Sedrakian2007}
 \bea\label{EMISSIVITY}
&&\epsilon_{\nu} =-\left( \frac{2 G}{\sqrt{2}}\right)^2
\int\!\frac{d^3k}{(2\pi)^3 2\omnu(\vec k)}
\int\!\frac{d^3k'}{(2\pi)^32 \ome(\vec k')}
\nonumber\\
&&\int\! d^4 q \,\delta (k + k' - q) g(q_0)\bar f_{e}(\ome)
\Img [\Lambda^{\mu\nu}(k,k')\,\Pi^R_{\mu\nu}(q)],\nonumber\\
\eea
where $k$ and $k'$ to the four-momenta of antineutrino and electron,
$\omega_\nu$ and $\omega_e$ are their on-mass-shell energies,
$\bar f(p) = 1-f(p)$ with $f(p)$ being the Fermi-Dirac distribution
function and index $e$ refers to electrons, 
$\Lambda^{\mu\nu} = {\rm Tr}\left[\gamma^{\mu} (1 - \gamma^5)\sla
  k\gamma^{\nu}(1-\gamma^5)\sla k'\right]$,
$g(q_0)$ is the Bose distribution function and $\Pi^R_{\mu\nu}(q)$
is the retarded polarization. We use the spectral representation of
the retarded Green's functions of neutrons and protons and compute the
polarization tensor in the one-loop approximation. After contracting
the leptonic and hadronic traces we obtain
\bea 
\label{eq:emissivity}
\epsilon_{\nu} &=& 2G^2 \int d^4q L_1(q) \, L_2(q),
\eea
where the loop integrals are given by 
\bea
\label{eq:I1}
L_1(q) &=&\int\!\! \frac{d^4p'd^4p}{(2\pi)^2} 
  \delta^{(4)}(p-p'+q) A_n(p)A_p(p')f(p')\bar f(p),\nonumber\\\\
\label{eq:I2}
L_2(q)&=&\int\!\! \frac{d^3k d^3k'}{(2\pi)^6} 
\bar f(k') \bar f(k) \omega_{\bar\nu}
  \delta^4(k+k'-q), 
\eea 
where $p'$ and $p$ refer to the four-momenta of the neutrons and
protons. The spectral functions of the nucleons with $\tau \in n,p$ are
given by
\bea
\label{eq:spectral_function}
A_{\tau}(p) = \frac{\gamma_{\tau}}{[\epsilon-\epsilon_{\tau}(p)]^2+\gamma_\tau^2/4},
\eea
where its width and quasiparticle energy are given by
$\gamma_{\tau} = -2 \Img \Sigma_{\tau}^R(p)$ and
$\epsilon_{\tau} (p) = \epsilon_{\tau}^{\rm kin}+\Re\Sigma_{\tau}^R(p)-\mu_{\tau}$,
respectively. Here $\Sigma_{\tau}^R(p)$ is the retarded self-energy, $\mu_{\tau}$ is
the chemical potential and $\epsilon_{\tau}^{\rm kin}(p)$ is the kinetic energy.
In the small $\gamma_{\tau}$ limit~\cite{Alm1995,Schnell1996}
\bea \label{eq:QP_spectral_function}
A_{\tau}(p) = 2\pi z_{\tau}\delta[\epsilon_{\tau}-\epsilon_{\tau}(p)]
-\frac{\gamma_\tau}{[\epsilon_\tau-\epsilon_{\tau}(p)]^2} + O(\gamma_{\tau}^2),\nonumber\\
\eea
where any integration over the second term is understood as Cauchy principal value,
$\epsilon_{\tau} (p) = p^2/2m_{\tau}^*-\mu_{\tau}$ is the quasiparticle spectrum, $m_{\tau}^*$ is
the quasiparticle mass that is determined by the on-shell derivative
of $\Re\Sigma_{\tau}^R(p)$ with respect to the momentum
\bea
\frac{m_{\tau}}{m_{\tau}^*} = 1+ \frac{m_{\tau}}{p_{F\tau}} 
\frac{\partial\Sigma_{\tau}^R(p)}{\partial p}\Bigg\vert_{p=p_{F\tau}},
\eea
$z_{\tau}$ is the wave-function renormalization, which we include in
the effective mass via $z_{\tau}m_{\tau}^*\to m_{\tau}^*$. The first
term in Eq.~\eqref{eq:QP_spectral_function} is the quasiparticle
spectral function corresponding to $\gamma_{\tau} =0$, whereas the
second term is the leading order contribution in small damping.  It
should be considered in combination with the uncertain fraction of the
nFL high-momentum tail of the fermion ensemble, which can additionally
affect our results below.

The Fermi golden rule expression for the emissivity can
be easily recovered in the limit $\gamma_{\tau} \to 0$. Indeed one
finds the standard expression by doing first integrations over the
energy components $p_0$ and $p_0'$ in Eq.~\eqref{eq:I1} using
Eq.~\eqref{eq:QP_spectral_function} and subsequently taking the
$q$ integral in Eq.~\eqref{eq:I1} using delta functions. Note that the
angle integrations eliminate the spatial parts of the four-vector
products in the matrix element and it can be approximated as
\bea
\sum\vert {\cal M}_{\rm Urca}\vert^2 = 64 (k\cdot p') (p\cdot
k')\simeq 64p_0p'_0k_0k'_0,
\eea
where the sum stands for the average over the initial and summation over
the final spins of baryons.  The $L_2$ integral is computed by
neglecting in the $\delta$ function the neutrino energy-momentum
compared to that of the electron, which fixes the momentum transfer to the
Fermi momentum of the electron. In this manner, we obtain the neutrino
emissivity
\bea
\label{eq:emiss_4}
\epsilon_{\nu}  &=&  G^2  \frac{p_{Fe}m^*_pm^*_n}{4\pi^5}
\int_0^{\infty}\!\! d\omega_{\nu}\omega_{\nu}^3 \nonumber\\
&\times &
\int_{-\infty}^{\infty}  d\omega   g(\omega) f_e(\omega_\nu-\omega) 
L_1(\omega, q =p_{Fe}).
\eea
The three-dimensional integral in Eq.~\eqref{eq:emiss_4} must be
computed numerically in general.  If the spectral function of protons
is taken in the limit $\gamma_p\to 0$ the integral can be computed
analytically and we find
\bea 
\label{eq:I_1:c}
L_1(q)=c T\ln \Bigg\vert \frac{1+\exp\left(-\frac{\epsilon_{\rm 
        min}-\mu_p}{T}\right)} {1+\exp\left(-\frac{\epsilon_{\rm 
        min}+\omega-\mu_p}{T}\right)}\Bigg\vert, 
\eea
where $c\equiv m^*_nm^*_p/2\pi q$ and 
 the minimal energy $\epsilon_{\rm min}$ arises from the limits
on the azimuthal angular integration and is given by
$$
\epsilon_{\rm min} = ({m^*_n}/{2q^2})
\left(\mu_n+\omega-\mu_p+{q^2}/{2m^*_p}\right)^2.
$$
Substituting Eq.~\eqref{eq:I_1:c} in Eq.~\eqref{eq:emiss_4} we obtain
the general form of nFL-Urca emissivity at finite temperatures for
$\gamma_\tau = 0$. Further simplifications arise at low temperatures, where
the small $\omega$ expansion leads to
\bea\label{eq:L1}
L_1(q)/cT\omega= \{\exp [(\epsilon_{\rm min }-\mu_p)/T]+1\}^{-1}
\eea
after
which the remaining integrals can be computed analytically. In the
limit $T\to 0$ one finds then
$L_1(q)/cT\omega=\theta (\mu_p-\epsilon_{\rm min })
=\theta ( \vert p_{Fe}+p_{Fp}-p_{Fn}\vert), $
 i.e., the quasiparticle result of Ref.~\cite{Lattimer1991} is
recovered from Eq.~\eqref{eq:emiss_4}.

We now turn to the finite width effects in the proton spectral
function, while keeping the neutron spectral function in the
quasiparticle approximation, given by the $\delta$ function term in
Eq.~\eqref{eq:QP_spectral_function}.  With the full spectral function
for protons given by Eq.~\eqref{eq:spectral_function} we find
\bea 
\label{eq:I_1:b}
L_1(q)= c\!\! \int\!\! d\epsilon
[f(\epsilon)-f(\omega + \epsilon)]
 \left\{\frac{1}{2}-\frac{1}{\pi}{\rm atan} 
\left[\frac{\epsilon_{\rm min}-\epsilon}{\gamma_p/2}\right]\right\}.\nonumber\\
\eea
For concrete calculations, we need to specify the spectrum of protons
and neutrons $\epsilon_\tau(p)$ and the width of the proton spectral
function. We adopt for neutrons and protons
$\epsilon_\tau (p) = p^2/2m_\tau^*-\mu_\tau$, where
$m_\tau^*/m_\tau =0.7$ and $\mu_\tau$ is taken to be equal to the Fermi
energy computed from the partial density $n_\tau= k_{F\tau}^3/3\pi^2$,
where $k_{F\tau}$ is the Fermi wave vector. We model the width of
protons by
\bea
\label{eq:gamma}
\gamma_p = a_p T^2 \left[1+\left(\frac{\epsilon-\mu_p }{\pi T}\right)^2\right],
\eea
where
$a_p = (\hbar v_{Fp} \mu_p^{-1}) \sum_{\tau=n,p} \bar \sigma_{p\tau}
n_\tau \mu_{\tau}^{-1} $, where $v_{Fp}$ is the proton Fermi velocity,
$n_\tau$ are the number densities and $\bar\sigma_{np} = 7$~fm$^2$ and
$\bar\sigma_{pp} = 4 $~fm$^2$ are the angle averaged cross sections
for neutron-proton and proton-proton scattering.  Here the factor
$ v_{Fp} \sum_{\tau=n,p} \bar \sigma_{p\tau} n_\tau $ is the
scattering frequency and $\mu_\tau^{-1}$ arises from the the
degeneracy factor $T/\mu_{\tau}$ per nucleon, which also leads to
$T^2$ scaling of $\gamma_p$ in Eq.~\eqref{eq:gamma}.  A linear in
temperature nFL term can appear in this equation, but we will not
consider it here.
\begin{figure}
\includegraphics[width=\linewidth]{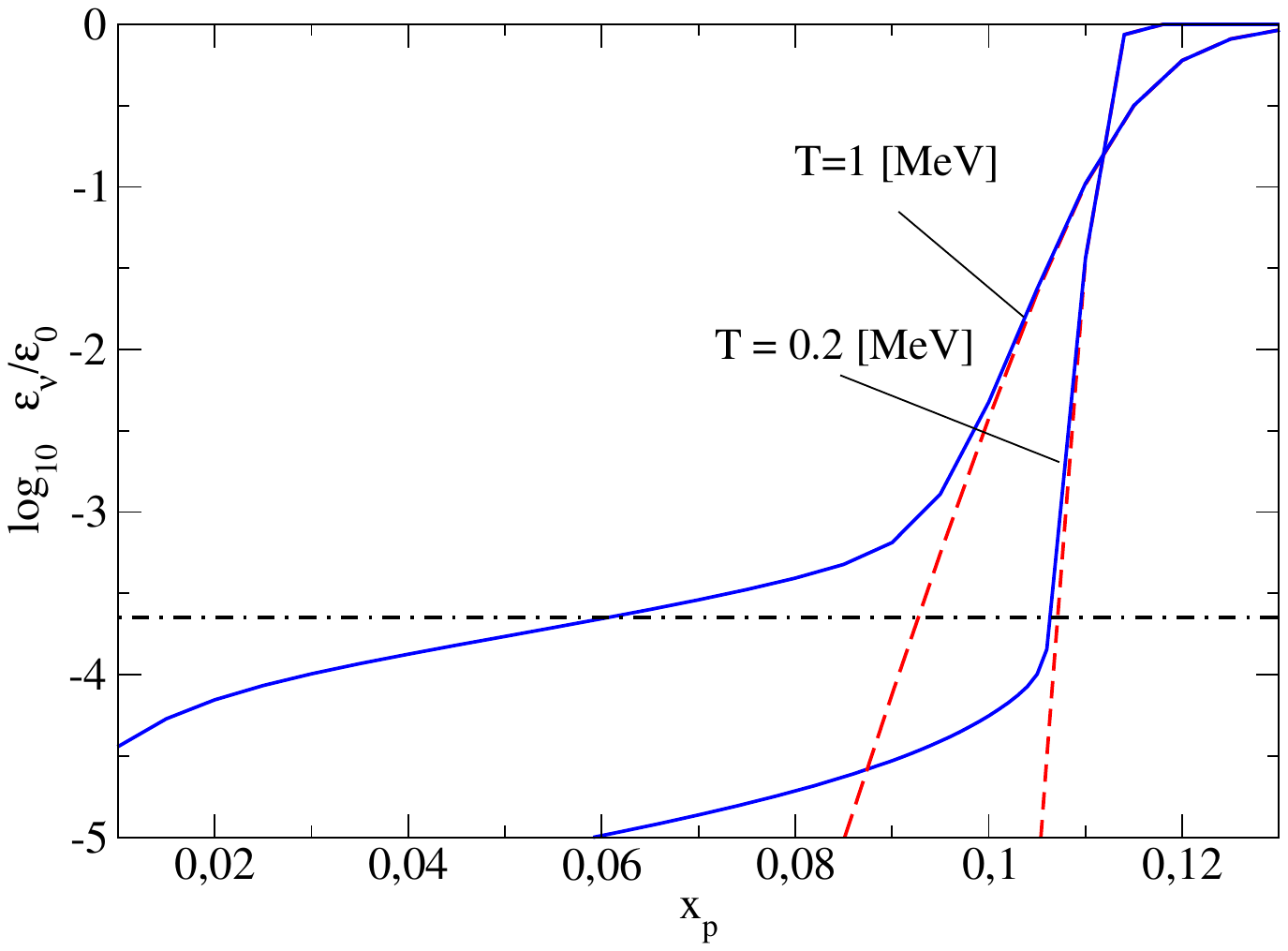}
\caption{\label{nFL-Urca_emiss_xp} The nFL-Urca emissivity in units of
  $\epsilon_0$ for $T = 1$ MeV and $T = 0.2$ MeV for density
  $\rho = 1.5\rho_0$ with $\rho_0=2.7\times 10^{14}$ g cm$^{-3}$ as a
  function of proton fraction $x_p$.  The dashed lines show the dUrca
  process, i.e., the quasiparticle limit corresponding to
  $\gamma_p = 0$.  For comparison, we show the FL one-pion-$\rho$
  exchange based mUrca process $\epsilon_{\rm mUrca}/\epsilon_0$ for
  $T = 1$~MeV by the dot-dashed horizontal line. The corresponding
  result for $T = 0.2$ MeV is $\epsilon_{\rm mUrca}/\epsilon_0=-5$.}
\end{figure}

{\it Results---}The neutrino emissivity was computed using
Eq.~\eqref{eq:emiss_4} and the spectral width is given by
Eq.~\eqref{eq:gamma}. In Fig.~\ref{nFL-Urca_emiss_xp} we show the
nFL-Urca rate normalized by the same quantity in the quasiparticle
approximation as a function of proton fraction $x_p=\rho_p/\rho$,
where $\rho_p$ and $\rho$ are the proton and total densities for two
fixed values of temperature $T=0.2$ and 1~MeV. In the quasiparticle
limit we observe the thermal ``activation'' of the dUrca process due
to the opening of the phase space when using the finite temperature
expression \eqref{eq:L1} instead of zero-temperature analog containing
$\theta$ function, as has been established long
ago~\cite{Yakovlev2001}.  We also show the $\pi$ plus $\rho$ meson
exchange interaction based result of Ref.~\cite{Friman1979} whose
dependence on the proton fraction cancels out by the normalization by
Eq.~\eqref{eq:epsilon_0}.  At the temperature 0.2 MeV
$\sim 2\times 10^{9}$~K the dUrca process is comparable to the mUrca
already for proton fraction $x_p=0.85$ instead of 0.11 in the
zero-temperature limit. Next, including the width of protons, which
implies $L_1$ given by \eqref{eq:I_1:b}, we see that the sharp
threshold is replaced by a ``smooth increase'' of the nFL-Urca rate as
the proton fraction increases. The fundamental difference between the
quasiparticle limit and the nFL-Urca process is that the last
dominates the mUrca processes even for small values of $x_p$ that are
compatible with the symmetry energy for the nuclear equation of
states, i.e., practically all the equations of states of neutron star
matter allow for the nFL-Urca process, if the width of protons induced
by the SRCs is included.
\begin{figure}
\includegraphics[width=\linewidth]{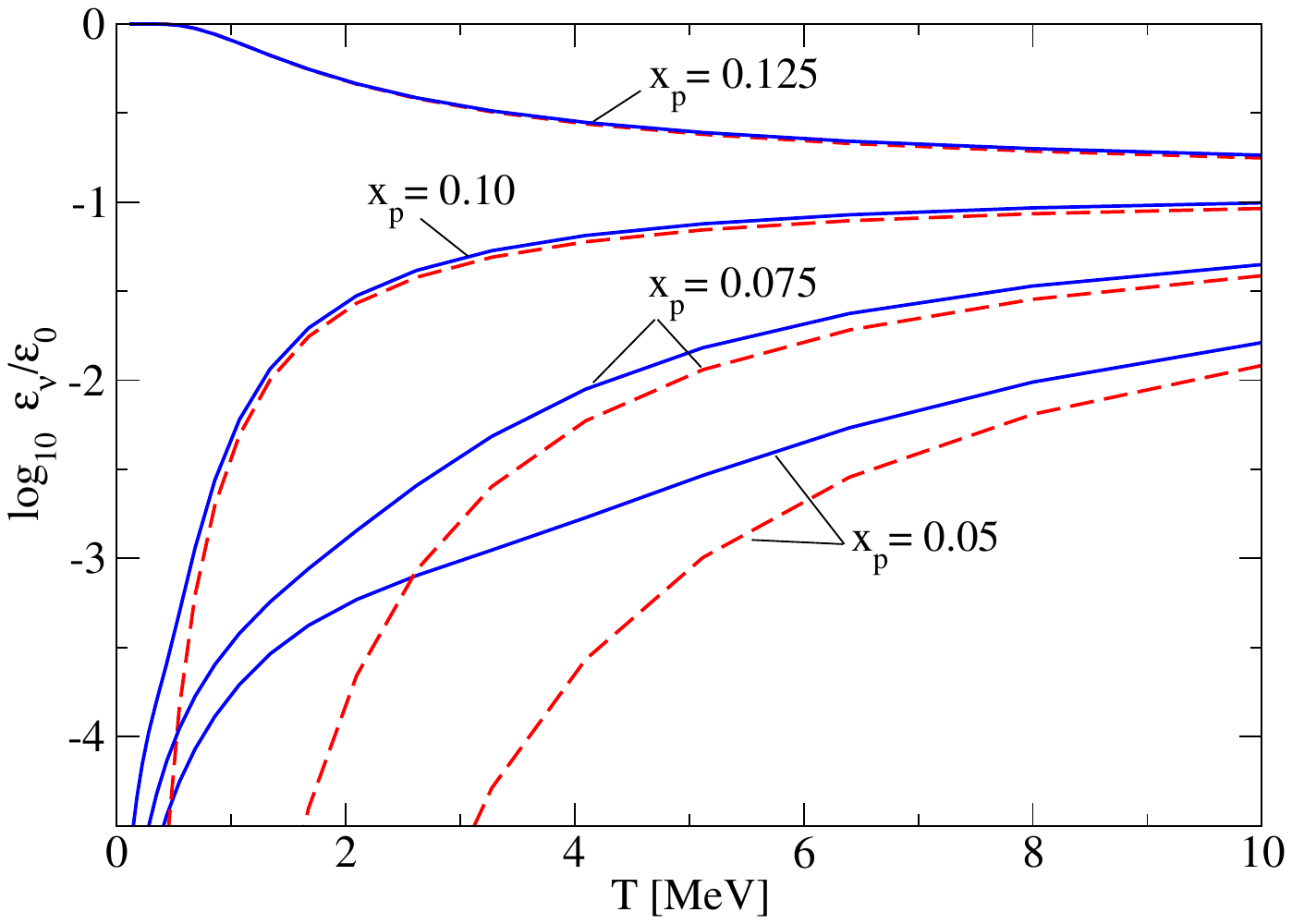}
\caption{\label{nFL-Urca_emiss_T} The nFL-Urca emissivity in units of
  $\epsilon_0$ for $x_p = 0.05$, 0.10 and 0.125 and $\rho = 1.5\rho_0$
  as a function of temperature.  The solid lines show the full result,
  whereas the dashed lines -- the limit $\gamma_p = 0$.  }
\end{figure}

Figure~\ref{nFL-Urca_emiss_T} illustrates the temperature dependence
of the nFL-Urca process for fixed proton fractions ($x_p$) and
compares it with the quasiparticle approximation of dUrca, but
including thermal activation.  Starting with the quasiparticle limit,
we observe the well-known behavior of the dUrca process: for
sufficiently large proton fractions (here $x_p = 0.125$), the dUrca
process is kinematically allowed and the finite width and quasiparticle
results nearly coincide.  In contrast, for lower proton fractions
(here $x_p = 0.1$, $0.075$, and $0.05$), the process is strongly
suppressed in the quasiparticle limit at low temperatures. However,
for temperatures around 1 MeV and higher, thermal activation of the
dUrca process becomes evident for these smaller values of $x_p$.
Including the proton's spectral width leads to significant differences
in the emissivity of nFL-Urca at low temperatures $T\le
1$~MeV. Instead of vanishingly small results, the emissivity decays
much more slowly.  Thus, it is evident that the finite width of
protons induced by SRCs smooths out the sharp
threshold on the dUrca processes in neutron star matter.

{\it Implications for neutron star cooling---}The common picture of
neutron star cooling assumes that for $x_p\ge 0.11$ the dUrca process
is operative, otherwise the mUrca process (which is permitted for any
value of $x_p$) is the dominant cooling mechanism. Such large values
of $x_p$ are achieved only in massive stars $M\ge 1.7~M_{\odot}$, the
exact value depending on the symmetry energy behavior of the
underlying equation of state at supranuclear densities.  As a
consequence, modern cooling simulations display a sharp drop in
temperatures of stars that have masses lying beyond the threshold mass
for the dUrca process. As a consequence the massive stars featuring
the dUrca process cool much faster than less massive stars that cool
by the mUrca process.

As we have hypothesized above, the inclusion of the proton spectral
width, which is one of the manifestations of SRCs in asymmetrical
nuclear matter leads to the nFL-Urca process, which is operative far
below the threshold proton fraction. Therefore, we may conjecture that
the cooling simulations including SRCs will smooth out the drop in the
stellar temperatures for sequences of stellar models crossing the mass
threshold for the opening of dUrca process.  It is well-known that
superfluidity suppresses direct Urca rates~\cite{Yakovlev2001}, and we
can anticipate a similar effect for the nFL-Urca rates obtained
here. However, it should be emphasized that, in addition to the
suppression from proton-proton pairing, supra-Fermi surface neutrons
and protons may form pairs~\cite{2014Sci...346..614H}  similar
to the Cooper pairing in the isospin singlet $^3S_1$-$^3D_1$
channel~\cite{Sedrakian2006,Stein2014}. This can potentially alter the
nature of the nFL contributions to the Urca rates below critical
temperature for Cooper pairing in this channel.

{\it Conclusions---}In conclusion, there is strong experimental
evidence for SRCs in asymmetrical nuclear systems,
manifesting as tails in the momentum distribution of protons and
off-shellness of their spectral width. These correlations have been
demonstrated in numerous experiments, particularly those conducted at
JLab~\cite{2014Sci...346..614H}. Consequently, the Fermi-liquid
approximation is likely invalid for protons in dense matter. In this
work, we obtained the emissivity of dense $n-p-e$ matter via the Urca
process, accounting for the finite width of protons. In the zero-width
approximation, we recover the pioneering result of
Ref.~\cite{Lattimer1991}. We have shown that the nFL-Urca process
remains effective even for low proton fractions, replacing the
threshold behavior of the dUrca process with a smooth dependence on
$x_p$.  Finally, it should be noted that from the perspective of
many-body nuclear theory, the distinction between the dUrca and mUrca
processes is strictly possible only in the quasiparticle
approximation. Once the width of the proton propagator is taken into
account, the higher-order interactions (higher-order loop diagrams)
are automatically taken into account. Therefore, nFL-Urca may be
viewed as an alternative many-body approach to account for the
response of neutron-proton mixture to weak perturbations. This has
been extensively discussed in the case of the neutrino bremsstrahlung
process~\cite{Raffelt1996,Raffelt1999,Sedrakian1999,Sedrakian2000}.

In the case of Urca processes, for example, the mUrca rate computed in
Ref.~\cite{Friman1979} can be viewed as the dUrca process with the
width computed in the second Born approximation to the nucleon
(neutron or proton) propagator in the one-pion (eventually plus
$\rho$-meson) exchange approximation. Nevertheless, the description of
Urca processes in terms of proton spectral functions allows one to
establish contact with the study of SRCs in nuclear systems, as well as
many-body formulations of nuclear theory that are based on spectral
functions. It also allows  us to trace the smooth physical dependence
of neutrino emission rates on the proton fraction, hence the
density dependence of the symmetry
energy of neutron star matter. This last point may have important
ramifications for the cooling studies of neutron stars.

Finally, it is important to note that SRCs can significantly affect
various other aspects of neutron star physics. These effects include
influencing pairing-type correlations~\cite{Sedrakian2019,Rios2017},
isospin-triplet proton polarons~\cite{Tajima2024PhLB}, transport
phenomena, and many others.

\vskip 0.5cm 

{\it Acknowledgements---}The author acknowledges support from Deutsche
Forschungsgemeinschaft Grant No. SE 1836/5-3 and the Polish NCN Grant
No. 2020/37/B/ST9/01937. Gratitude is extended to the Institute for
Nuclear Theory at the University of Washington for its kind
hospitality, stimulating research environment, and partial support
through the INT's U.S. Department of Energy Grant
No. DE-FG02-00ER41132 during INT workshop ``Astrophysical neutrinos
and the origin of the elements.''  The author also gratefully
acknowledges discussions on related topics with M. Alford, A. Haber,
A. Harutyunyan, and S. Reddy.

\newcommand{\physrep}{Physics Reports}

\providecommand{\href}[2]{#2}\begingroup\raggedright\endgroup

\end{document}